\documentclass[aps,pre,showpacs,showkeys,preprintnumbers,%
amssymb,amsfonts,floatfix,a4paper,twoside,twocolumn]%
{revtex4}

\usepackage[pdftex]{graphicx}
\DeclareGraphicsExtensions{jpg,tif,pdf}

\usepackage{amsmath}
\usepackage{bm}
\usepackage{time}
\usepackage{fancyhdr}
\usepackage{float}

\renewcommand{\vec}[1]{\mathbf{#1}}

\newcommand{\ie}{{\it i.e.}, }

\newcommand{\eg}{{\it e.g.}, }

\newcommand{\beq}{\begin{equation}}
\newcommand{\eeq}{\end{equation}}
\newcommand{\bega}{\begin{eqnarray}}
\newcommand{\ega}{\end{eqnarray}}

\usepackage{color}
\usepackage{soul}
\usepackage{ulem}
\definecolor{lightgray}{cmyk}{0,0,0,.1}

\sethlcolor{lightgray}

\begin{document}

\title{Frequency dependent deformation of liquid crystal droplets in an external
  electric field}

\date{\today -- \now}

\author{G\"unter K. Auernhammer}
\email{auhammer@mpip-mainz.mpg.de}

\author{Jinyu Zhao}

\author{Beate Ullrich}

\author{Doris Vollmer}

\affiliation{
  Max-Planck-Institut f\"ur Polymerforschung,
  Ackermannweg 10,
  55128 Mainz,
  Germany
}

\begin{abstract}
  Nematic droplets suspended in the isotropic phase of the same
  substance were subjected to   alternating electrical fields of
  varying frequency.
  To keep the system at a constant nematic/isotropic volume ratio
  with constant droplet size we carefully kept the temperature in
  the isotropic/nematic coexistence region, which was broadened
  by adding small amounts of a non-mesogenic liquid. 
  Whereas the nematic droplets remained spherical
  at low (in the order of 10 Hz) and high frequencies (in the order of
  1 kHz),
  at intermediate frequencies we observed a marked flattening of
  the droplets in the plane perpendicular to the  applied field.
  Droplet deformation occurred both
  in liquid crystals (LCs) with positive and negative dielectric
  anisotropy. 
  The experimental data can be quantitatively modelled with a
  combination of the leaky dielectric model and screening of the
  applied electric field due to finite conductivity.
\end{abstract}

\keywords{}
\maketitle

\pagestyle{fancy} \fancyhead[RE]{} %
\fancyhead[LE]{\slshape G. K. Auernhammer, et al.}%
\fancyhead[RO]{\slshape G. K. Auernhammer, et al.: LC droplet
deformation in E-field}%
\fancyhead[LO]{}

\section{Introduction}
\label{sec:introduction}


Liquid crystal droplets dispersed in isotropic media have attracted
strong interest. Driven by new applications, during the last few years
the research focus has shifted from studying 
the director field and defect structure in quasi-static situations
(see, \eg \cite{crawford1996,candau1973,doane1986,chan1999,lubensky1998,%
dolganov2007}) to dynamic processes under the influence of
external fields \cite{park1994,lev2001,wu2007,elsadek2007}. Due to
the anisotropic nature of the liquid crystalline phases, these
droplets show a much richer physics than purely isotropic systems.

The literature on the behavior of liquid crystalline droplets
subjected to external fields is dominated by studies on droplets
suspended in a non-mesogenic carrier fluid. Owing to the
increasing importance of polymer dispersed liquid crystals in
various applications, the scientific interest in the response of
droplets to external field has grown significantly (for reviews
see, \eg \cite{crawford1996,coates1995,bouteiller1996,bunning2000,%
tjong2003,jeon2007}). In addition, magnetic \cite{candau1973} and
shear fields \cite{wu2007} have been applied to these systems. 
Only few studies have been carried out on liquid crystals in the
coexistence region. 
Park and coworkers \cite{park1994} studied nematic droplets in
a quasi 2-dimensional system submitted to an electric field
superimposed with a temperature gradient. Dolganov and coworkers
\cite{dolganov2007} investigated nematic droplets confined by a
surrounding smectic phase.

Inevitable impurities in liquid crystal samples are known to open
up a coexistence region between the nematic and
isotropic phase (see, \eg \cite{chiu1995,shen1995,lin1998,%
dasgupta2001,mukherjee2002}). 
This coexistence region allows to
have nematic droplets immersed in the isotropic phase of the same
liquid at constant sample temperature.
The physics of droplets in the isotropic-nematic coexistence
region is peculiar because, unlike typical liquid-liquid
interfaces, the interfacial tension is very low. Since droplet
deformation always implies an increase in surface area, the higher
the surface tension, the more rigid the droplets appear. In contrast,
droplets in the nematic-isotropic coexistence region have a surface
tension \cite{langevin1973,faetti1984,williams1976,yokoyama1985} of
the order of $10^{-5}$ Nm and can be deformed very easily.

In this paper, we discuss electric field-induced droplet deformations
that are peculiar to systems with low interfacial tension. 
We put special emphasis on the influence of field
frequency, and the actual electric field strength. 
Moreover, we took special care to avoid electrochemical reactions
in the sample that might influence the results. 
%
The paper is organized as follows:
In Sec.~\ref{sec:experimental-details} we describe the materials
and the experimental set-up. 
In Sec.~\ref{sec:colloid-free-system}, we present the results on the
droplet deformation in the colloid-free systems.
Sec.~\ref{sec:tracing-flow-field} is devoted to investigating the
underlying hydrodynamic flow field by adding tracer colloids to the
system.  We then compare our
results to theoretical predictions in Sec.~\ref{sec:theor-cons}.

\section{Experimental details}
\label{sec:experimental-details}

\begin{figure}
  \centering
  \includegraphics[width=0.3\textwidth]{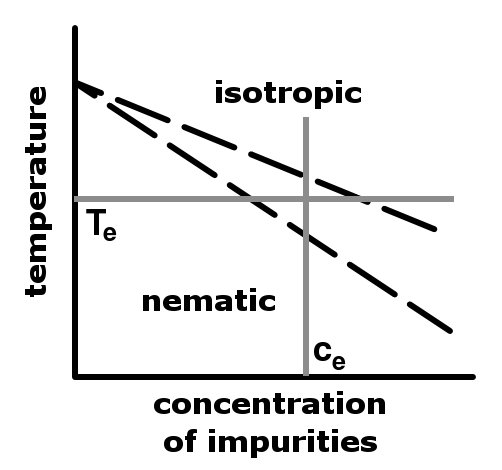}
  \caption{Schematic phase diagram of a liquid crystal with added
    impurities (in this work: octane). With increasing octane
    concentration, 
    the nematic isotropic phase transition shifts to lower
    temperatures and a nematic-isotropic coexistence region opens.
    The temperature $T_e$ and the octane concentration $c_e$ (gray
    lines) were chosen be well within the coexistence region.}
  \label{fig:phase-dia}
\end{figure}

Both the electric and dielectric properties of nematic liquid crystals
are markedly anisotropic.
%
%
Whereas
the conductivity $\sigma$ typically exhibits positive
anisotropy
\begin{equation}
\sigma_{\parallel} > \sigma_{\perp}
\end{equation}
(the indices $\parallel$ and $\perp$ indicating the conductivity
parallel and perpendicular to the nematic director $\hat n$,
respectively), the anisotropy of the dielectric constant
$\epsilon$  can vary (and change sign) between different
substances
\begin{equation}
\epsilon_{\parallel} \lessgtr \epsilon_{\perp}.
\end{equation}
We chose a typical representative for either class:
methyloxybenzylidenbutylaniline (MBBA, Frinton Laboratories, USA,
$\Delta \epsilon < 0$) and pentylcyanobiphenyl (5CB, Chemos,
Germany, $\Delta \epsilon > 0$). Both chemicals were used as
supplied. Impurities like polymers or low molecular weight organic
sovents in liquid crystals lead to coexistence of the
nematic and the isotropic phase within a certain temperature interval
\cite{shen1995,lin1998,roussel2000,ullrich2006} (see
Fig.\ref{fig:phase-dia} for a schematic representation of the
phase diagram).
In our experiments, adding octane (Aldrich, analytic grade) at a
concentration of \mbox{$c_e = 2$ vol\%} resulted in a coexistence
interval of the order of \mbox{1.5 K}, which is
%
sufficient to override the effect of any
impurities inevitably present in commercially available liquid
crystals. The results presented in this paper did not change for
octane concentrations between 1 and \mbox{4 vol\%}.
Poly(methyl methacrylate) particles (PMMA) with a radius of
approximately \mbox{500 nm} sterically stabilized by chemically
grafted poly(12-hydroxy-stearic acid) molecules
\cite{antl1986,bosma2002} were used as tracer particles. 

\begin{figure}
  \centering
  \includegraphics[width=0.5\textwidth]{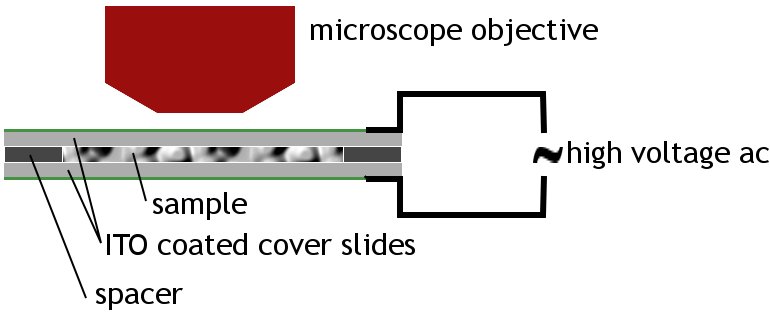}
  \caption{The sample was placed between two ITO coated
    glass slides (purchased form SPI Supplies) without direct contact between
    the sample and the electrodes. The sample thickness is approximately 180
    $\mu$m and the electrode spacing 500 $\mu$m. After loading, the
    sample was sealed with UV glue (Norland Products, USA). Within the
    experimental time, typically less than 5 hours, no changes in the
    properties of the sample were observed.
    }
  \label{fig:set-up}
\end{figure}

As depicted in Fig.~\ref{fig:set-up}, the sample was placed
between two glass slides of $\approx 170$ $\mu$m each. The outer
side of these slides was covered with transparent indium-tin-oxide
(ITO) electrodes. This configuration ensured that there was no
electric contact between the electrodes and the sample. The
surface of the glass slides was otherwise untreated. The sample
cell was temperature controlled to a precision better than 0.05 K
using a Linkam heating stage (THMS 600). All observations were
performed using an Olympus BX51 optical microscope with a long
working distance 20x objective and equipped with a CCD camera. For
better contrast and good resolution of droplet edges, the sample was
placed between parallel polarizers. The ITO electrodes allowed to
subject the sample to an alternating electric field parallel to
the light path of up to \mbox{0.4 V/$\mu$m} and with frequencies
up to \mbox{5 kHz}. An oscilloscope was used to ensure
that the applied voltage remained constant over the whole
frequency range.

After placing the sample in the measurement cell we first
thermally equilibrated the system for 20 to 30 minutes at a
temperature several degrees above the isotropic/nematic phase
transition temperature $T_{IN}$. Then we
slowly cooled the system into the coexistence region (typically 0.1 
to \mbox{0.2 K} below $T_{IN}$). 
Until the diameter of the nematic droplets formed in the sample
reached 50 to 150 $\mu$m. 
The sample was then
equilibrated at this temperature for 10 to 15 minutes. In this
way, the temperature could attain a constant value
throughout the sample and the octane concentrations in both phases
could adjust to the values given by the phase diagram. Since the
density of nematic droplets is a few \mbox{$10^{-3}$ g/cm$^3$}
greater than that of the isotropic matrix (see
\cite{deschamps2008} and references therein), they sink to the
bottom of the cell during this second relaxation time.

Three types of frequency dependent measurements were performed: i)
logarithmic increase of the frequency with time from low to high
frequencies, \ie constant time per frequency decade, ii)
logarithmic decrease from high to low frequencies, and iii)
switching directly from zero field to the desired frequency and
field strength. In all three cases, the two-dimensional projection
of the droplet size onto the observation plane was measured as a
function of time using ImageProPlus software (Media
Cybernetics, USA).

\begin{figure*}[tb]
  \includegraphics[width=0.8\textwidth]{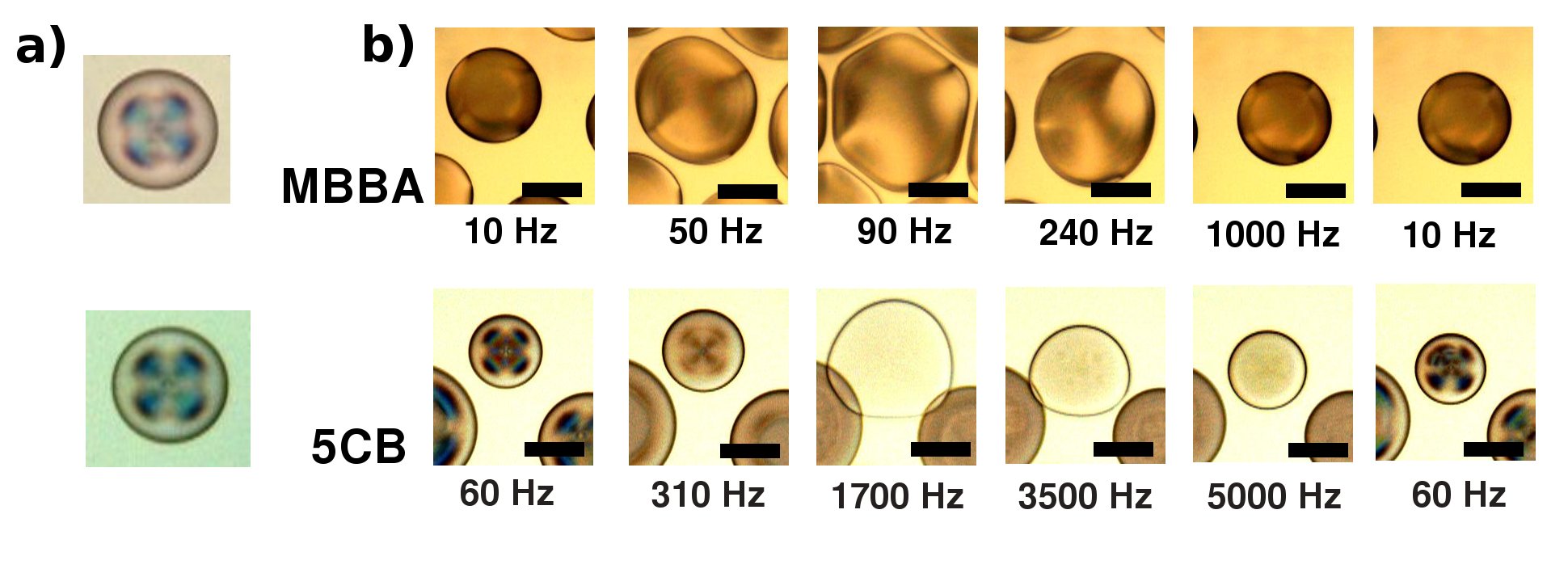}
  \caption{a) Images of nematic droplets of MBBA and 5CB dispersed
    in the isotropic phase in the absence of an applied electric
    field.
    b) In the series of images taken during frequency sweeps (sweep
    time: \mbox{100 s} for MBBA: \mbox{10 Hz} to \mbox{1 kHz} and
    for 5CB: \mbox{50 Hz} to \mbox{5 kHz}, voltage \mbox{0.4
    V/$\mu$m}) the droplet shape and apparent size changes as a function
    of frequency. The scaling bar indicates \mbox{100 $\mu$m}. }
  \label{fig:series}
\end{figure*}

\section{Results}
\label{sec:results}
\subsection{Colloid-free system}
\label{sec:colloid-free-system}

Figure \ref{fig:series} shows the results of typical frequency
sweeps in MBBA and 5CB. Part a) shows field-free droplets, whereas
in part b) all images were taken at 0.4 V/$\mu$m. Now, the
director orientation in the droplets can be assumed to be
predominantly perpendicular (MBBA) or parallel (5CB) to the
optical path. In the absence of any applied field, there was no
significant difference between MBBA and 5CB droplets. The observed
texture of the droplets was consistent with spherically symmetric
director orientation inside the droplets \cite{ondris1991}.

When an electric field was applied, the texture of the MBBA
droplets changed, as shown in Fig.~\ref{fig:series}. This was
caused by the negative dielectric contrast $\Delta \epsilon$ of
MBBA which leads to a predominantly in-plane director orientation.
In the case of 5CB, the texture hardly change when
turning on a low frequency field. This might be surprising
since many of the material parameters (like the elastic constants)
of the two samples are similar (see
Tab.~\ref{tab:mater-param}). In 5CB, however, the applied field
did not change the symmetry along the optical path. Turning the
director inside of the droplet parallel to the field and to the
optical path keeps the rotational symmetry of the droplet
projection through the microscope. This is different for MBBA,
where the electric field turns the director perpendicular to the
optical path and thus breaks the rotational symmetry of the
projection image of the droplet.

\begin{table*}
\begin{tabular}{|lc|cc|cc|}
\hline
 && \multicolumn{2}{|c|}{\bf MBBA} & \multicolumn{2}{|c|}{\bf 5CB} \\
\bf Quantity && Value & Ref. & Value& Ref. \\
\hline
Interfacial tension & $\gamma$
	& $1.5\cdot10^{-5}$ Nm$^{-1}$ & \cite{williams1976}
	& $1.6\cdot10^{-5}$ Nm$^{-1}$ & \cite{faetti1984} \\
	\hline
Dielectric constant & $\epsilon_\parallel$
	& 4.85 & \cite{diguet1970} & 15.5 & \cite{chandra77} \\
	& $\epsilon_\perp$ & 5.1 & \cite{diguet1970} & 7.5 & \cite{chandra77} \\
	& $\epsilon_{iso}$ & 5 & \cite{degennes93} & 10.5 & \cite{chandra77}\\
Dielectric ratio & $S=\frac{\epsilon_m}{\epsilon_d}$ & 0.92 & & 0.65 &\\
	\hline
Electric conductivity & $\sigma_m$
	& $\mathit{5\cdot10^{-8}}$ Sm$^{-1}$ 
	&& $\mathit{1.5\cdot10^{-6}}$ Sm$^{-1}$&\\
Conductivity ratio & $ R = \frac{\sigma_d}{\sigma_m} $ & 0.9 
	& \cite{diguet1970} & 1.1 & \cite{jadzyn1987} \\
	\hline 
Viscosity ratio & $M = \frac{\nu_m}{\nu_d}$ & 2.1 & 
	\cite{park1994} & 0.7 & \cite{park1994} \\
	\hline 
Elastic constant & $K_1$ & $6 \cdot 10^{-12}$ N &
	\cite{haller1972} & $6.2 \cdot 10^{-12}$ N & \cite{stewart2004} \\
	& $K_2$ & $3.8 \cdot 10^{-12}$ N & \cite{haller1972} & $3.9 \cdot
	10^{-12}$ N & \cite{stewart2004} \\
	& $K_3$ & $7.5 \cdot 10^{-12}$ N & \cite{haller1972} & $8.2 \cdot
	10^{-12}$ N & \cite{stewart2004} \\
	\hline	
\end{tabular}
\caption{Material parameters of MBBA and 5CB in the coexistence
region. The electric conductivities 
were fitted to match the model with the experimental data as shown
in Fig.~\ref{fig:mbba-deform-comp}. Still these values come within the
limits typically expected for these materials.
Note that $S$ is calculated for oriented droplets in which the
director is perpendicular (MBBA) or parallel (5CB) to the applied
field, with $\epsilon_m$ and $\epsilon_d$ being the dielectric
constants of the continuous medium and the droplets, respectively. 
}
\label{tab:mater-param}
\end{table*}


Changing the frequency of the electric field not only changed the
apparent size of the droplets, but also their internal texture,
indicating a change of director orientation. In 5CB, this effect
increased with frequency, corresponding to a good director orientation
parallel to the field. In MBBA, the change was not visible due to the
negative dielectric constant. From the evolution of the texture we can
conclude that the higher the frequency,the better the orientation
inside the droplets, \ie the higher the electric field inside the
sample. 

All commercial liquid crystals have a small but finite conductivity
based on ionic impurities. Since the electrodes are electrically
insulated from the sample, 
we expect that the field in the sample is smaller than the applied
electric field due to screening by ion migration. At low
frequencies, the ions can easily follow the applied field and screen 
the sample, whereas at sufficiently high frequency, the limited
ion mobility prevents this.
Thus, the field is more markedly reduced at low than at high
frequencies. 
The increase of the field as a function of frequency is responsible
for the increasing orientation of the nematic droplets in 5CB as shown
in Fig.~\ref{fig:series}b). In Sect.~\ref{sec:screening} we shall
give a quantitative description of this effect.

\begin{figure}
  \centering
  \includegraphics[width=0.45\textwidth]{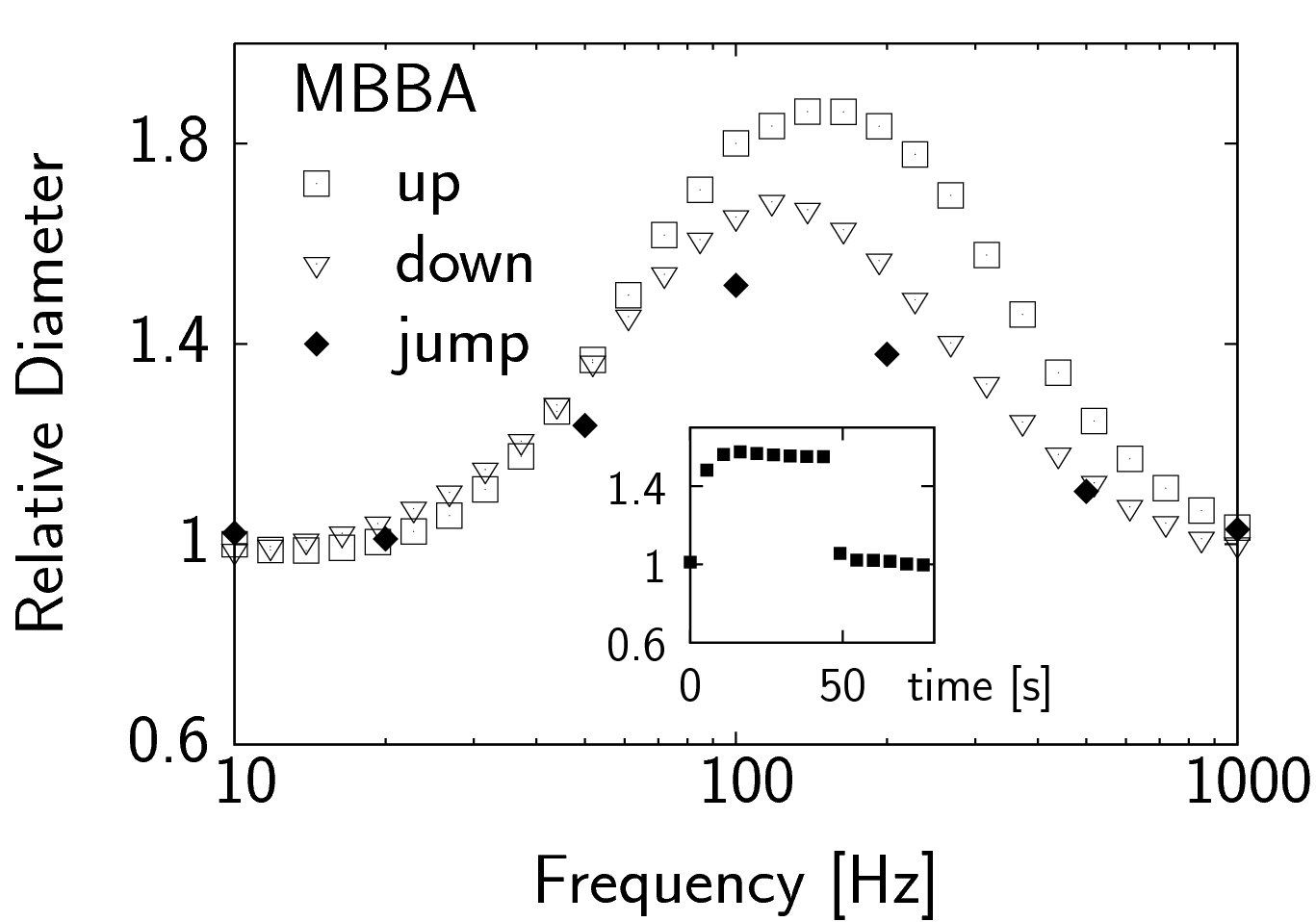}
  \caption{The typical evolution of MBBA droplets  while changing
    the frequency in three different ways (up: logarithmic sweep from
    10 Hz to 1 kHz; down: logarithmic sweep from 1 kHz to 10 
    Hz; jump: each frequency is directly addressed from a field free state).
    The apparent droplet diameters are normalized by their diameter before
    applying the field, which were measured to be 135
    $\mu$m (up), 114 $\mu$m (down), and 91 $\mu$m (jump). The
    different amplitudes of the curves  are due to the different
    initial droplet size. The sweeps in the major plot were
    performed with \mbox{100 s} sweep time. The inset shows the time
    dependence of the relative diameter after turning on the voltage
    (jump at 100 Hz).
  }
  \label{fig:mbba-amplitude}
\end{figure}

As can be seen in Fig.~\ref{fig:series}b) (5CB) and in
Fig.~\ref{fig:mbba-amplitude} (MBBA), the amplitude of the
diameter change depends on the initial droplet size.
In both 5CB and MBBA, droplets with an initial
(field-free) diameter of 50 -- 200 $\mu$m exhibited marked changes
of the apparent size after field application.
In MBBA, the amplitude ratio, \ie the ratio of the maximum diameter to
the initial diameter,  varied from  1.2 for droplets initially
50 $\mu$m in diameter to values up to 2 for initial diameters of 140
$\mu$m and decreased again for even larger droplets. 
A similar behavior was observed in 5CB. Note that
the frequency for the maximum diameter in 5CB is higher by about
one order magnitude than in MBBA (see also Sect.~\ref{sec:theor-cons}). 

Despite all these differences, the observed behavior was very similar
in both systems. 
Furthermore, the samples could be used in many frequency sweeps (at
least for several hours) without any noticeable change in the
frequency response. 
The electrical insulation between the electrodes and the sample
suppresses electrochemistry as a possible mechanism of degradation. 
However, the sample properties did change slightly in
samples stored for more than 24 hours. 
For that reason, we only consider data taken within the first 5 hours
after sample preparation. 
This virtually excludes field-induced damage as has been reported in
experiments on electro-hydrodynamic convection (see, \eg
\cite{richter1995}). 

The shape change was almost independent of frequency history as
shown for MBBA in Fig.~\ref{fig:mbba-amplitude}. The amplitude
variations are in the range expected for droplets of different size. 
However, the frequency of maximum
amplitude $f_{max}$ was slightly less in the case of decreasing
frequency sweeps. This indicates a small memory effect which might
arise from ohmic heating in the ITO layers or from relaxation
phenomena as shown in the inset to
Fig.~\ref{fig:mbba-amplitude}. After a jump in the  
applied voltage it takes 10 -- 15 s until the droplet has attained
its stable shape. 

As temperature is kept constant, also the nematic volume fraction and
droplet volume can be considered constant. 
%
Therefore, the observed increase in apparent diameter must be caused
by flattening of the droplets, 
\ie the droplets take an oblate (disk-like) shape with  symmetry
axis  along the electric field (as sketched in
Fig.~\ref{fig:mbba-amplitude}). 
This behaviour differs from that predicted by purely dielectric
considerations 
\cite{taylor1964}. 

\subsection{Tracing the flow field}
\label{sec:tracing-flow-field}

To test whether droplet deformation is of hydrodynamic origin, we
added tracer colloids to the sample mixture prior to filling the
chamber. In the coexistence region, these colloids remain exclusively in
the isotropic part of the sample. Thus they visualize the flow field
around the droplets. 

\begin{figure}
\centering
\includegraphics[width=0.5\columnwidth]{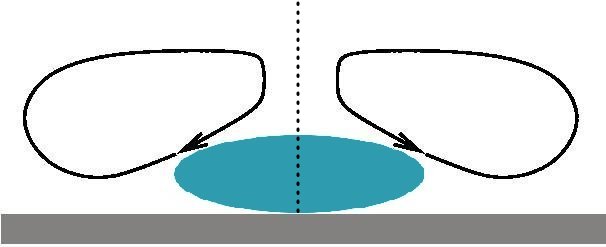}
\caption{Sketch of the flow field around the nematic droplets as
observed from tracer colloids. The flow field is axisymmetric around
the center axis of the droplet (dotted line).} 
\label{fig:flow-field}
\end{figure}

These experiments revealed a strong flow within the isotropic phase
around the nematic droplets. 
The tracer colloids moved outwards close to the
droplet surface, turned upwards well inside the isotropic areas,
returned above the droplet and then closed their convection role by
moving downwards from above the droplet to its side at some distance
from the interface. The resulting toroidal flow field is sketched in
Fig.~\ref{fig:flow-field}.  
The projected velocity of the tracer colloids was highest close to
the surface of the droplet. This indicates that the flow is driven
by the interface (see Sect.\ref{sec:theor-cons} for more details).
Via viscous friction, the toroidal flow field leads to a net outward
force on the droplet which leads to the 
transition from the spherical to the oblate droplet shape.

\begin{figure}
\centering
\includegraphics[width=\columnwidth]{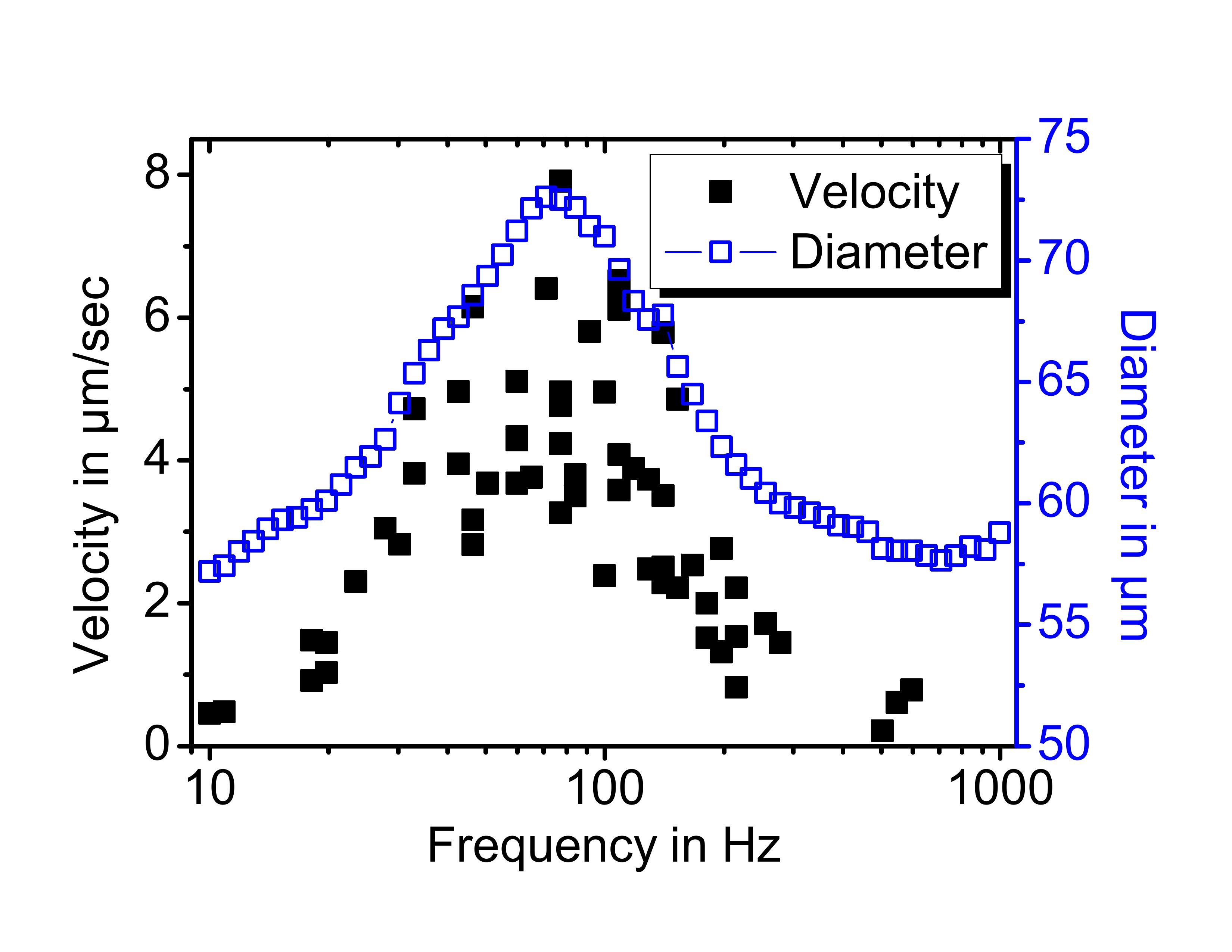}
 \caption{Characteristic flow velocities around a droplet (filled
   squares) and the  corresponding apparent diameter of the droplet
   (open squares). To have a velocity measure which is independent
   from the actual droplet size, all flow velocities are measured at
   the perimeter of the maximum droplet size, \ie for this droplet of
   a maximum diameter of \mbox{75 $\mu$m} at about \mbox{40 $\mu$m}
   from the droplet center. 
 }
 \label{fig:velo-tracer}
\end{figure}

The velocity maximum at intermediate frequencies coincides with
maximum deformation (Fig.\ref{fig:velo-tracer}). 
Within the accuracy of the data, the flow velocity vanishes at high
and low applied frequencies. 
Flow fields around droplets also lead to long-range interaction
between the droplets \cite{baygents1998}. In our low surface tension
system, this interaction leads to deviations from the circular shape,
as shown in Fig.~\ref{fig:series}b). 

\section{Theoretical considerations}
\label{sec:theor-cons}

\subsection{Screening}
\label{sec:screening}

\begin{figure}
\centering
\includegraphics[width=0.48\textwidth]{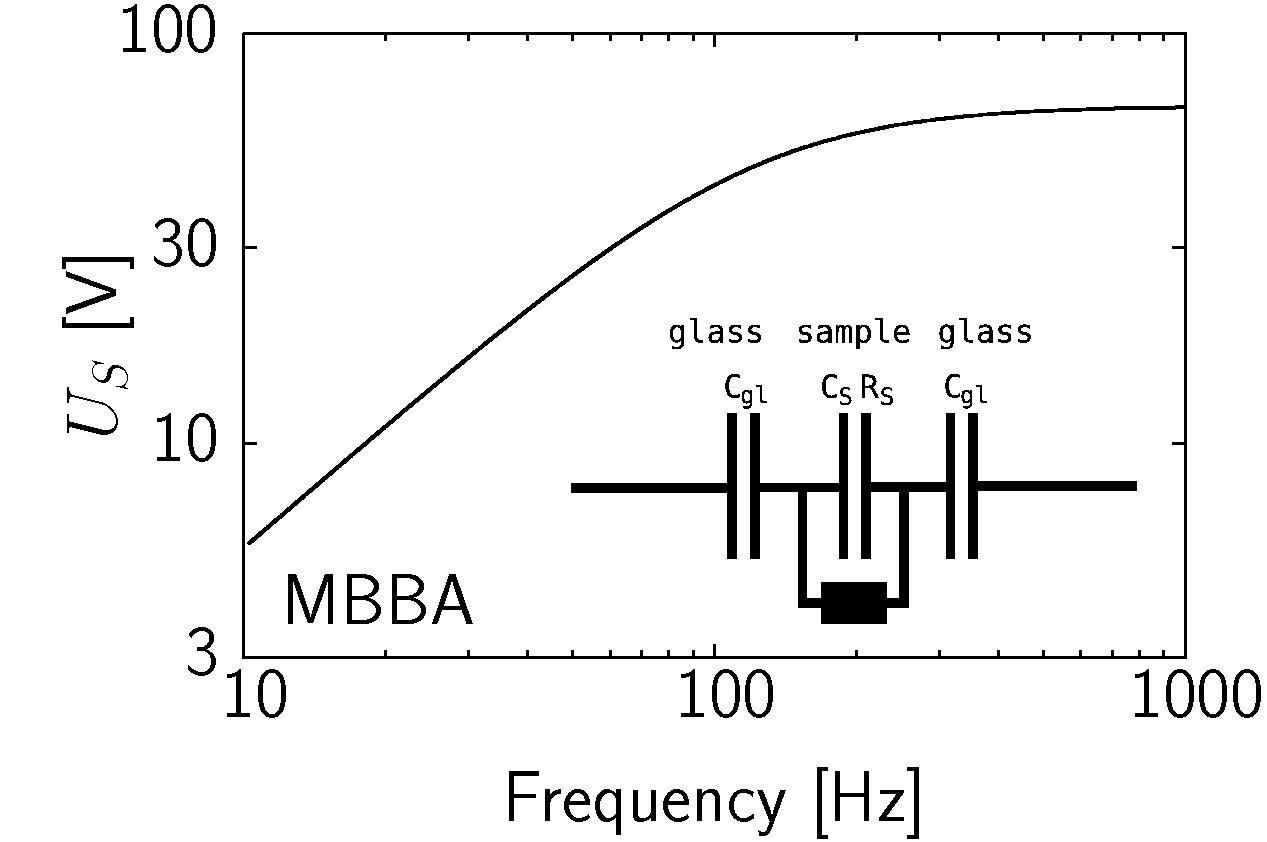}
\caption{Modeling the effects of a small but finite conductivity
of the samples. For typical values for MBBA ($\epsilon_s = 5$,
$\sigma = \sigma_m= 0.5\cdot 10^{-7} S/m$, $\epsilon_{gl} = 5$, and 
$U_0 = 200 V$)
the actual field in the sample reaches only at a few hundred Hz its
purely dielectric value. }
\label{fig:ldm-estimates}
\end{figure}

To describe the screening of the electric field due to the finite
conductivity of the sample, we use the model circuit sketched in the
inset in Fig.~\ref{fig:ldm-estimates}. Here, we model the
glass slides as pure capacitances, $C_{gl}$, and the sample as a
capacitance, $C_s$, in parallel with an resistance, $R_s$ (assumed to
be ohmic). The impedance ($Z_{cell}$) of this model circuit is given
by the sum of the impedances of the glass slides, $Z_{gl}$, and the
sample, $Z_s$ 
\begin{eqnarray}
Z_{gl} & = &-\frac{i}{\omega C_{gl}} \\
Z_s & = & (\frac{1}{R_s} + i \omega C_s)^{-1}\\
Z_{cell} & = & 2 Z_{gl} + Z_s,
\end{eqnarray}
with $i = \sqrt{-1}$, $\omega$ is the angular frequency of the applied
field, the capacities given by $C = (\epsilon_0 \epsilon A/d)$, the
ohmic resistance $R_s  =  d/(A \sigma)$, the
sample area by $A$, and the thickness $d = d_{gl} = d_s$. The 
voltage, $U_s$, actually applied to the sample given by 
\begin{eqnarray}
U_s  & = & U_0 \frac{Z_s}{Z_{cell}} \nonumber \\
& = & U_0  \frac{\omega \epsilon_0 \epsilon_{gl}}
        {2\omega \epsilon_0 \epsilon_s - 2i\sigma
         + \omega \epsilon_0 \epsilon_{gl}},
\label{eq:actual-field-sample}
\end{eqnarray}
where $\epsilon_0 = 8.85\cdot 10^{-12} F/m$ is the permittivity of
vacuum, $\epsilon_s$ and $\epsilon_{gl}$ are the dielectric constants
of the glass slides and the sample, and $\sigma$ is the conductivity
of the sample. Taking as typical values for MBBA $\epsilon_s = 5$,
$\sigma = 0.5\cdot 10^{-7} S/m$, $\epsilon_{gl} = 5$, and $U_0 = 200 V$, we
found that the voltage actually applied to the sample, $U_s$, reaches
the value expected from purely dielectric considerations,
$U_s^{dielect.}$ only for frequencies of a few hundred Hz (see also
Fig.~\ref{fig:ldm-estimates})
\begin{equation}
U_s^{dielect.} = 
	U_0 \frac{\epsilon_{gl}}{2\epsilon_s+\epsilon_{gl}} =
	\frac{U_0}{3} = \lim_{\frac{\sigma}{\omega} \to 0} U_s.
\end{equation}
As expected, the sample is essentially field-free at low frequencies
\begin{equation}
 \lim_{\omega \to 0} U_s = 0.
\end{equation}

\subsection{Leaky dielectric model}
\label{sec:leaky-diel-model}

The ``leaky
dielectric model'' describes the behavior of a droplet immersed in an
immiscible medium when subjected to an electric field. In this model,
both the droplet and the continuous phase are assumed to be 
slightly conducting dielectrics. The foundations of that model
have been laid by Taylor \cite{taylor1966}, Torza \cite{torza1971},
Saville \cite{vizika1992,saville1997}, and their coworkers. Recently
this model has experienced more detailed experimental
\cite{bentenitis2005,baygents1998} and numerical
\cite{fernandez2005,zhang2005} verifications.

Here, we include a brief review of the leaky dielectric model (see the
references cited above for details).
Any electric current $\vec j$ induced by an applied field has to
fulfill the continuity equation $\nabla \cdot \vec j =
0$. Additionally, the electric field is submitted to the usual
boundary conditions at the interface between the droplet and the
medium $\nabla \cdot \epsilon \epsilon_0 \vec E = \rho_e$ and $\nabla
\times \vec E = 0$. The combination of these conditions typically
leads to a finite electrical charge density on the droplet surface.  
\begin{figure}
\centering
\includegraphics[width=0.35\textwidth]{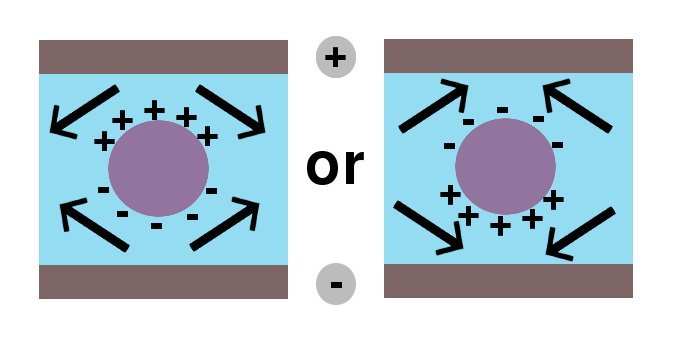}
\caption{The combination of electric boundary conditions and
continuity equation lead to surface charges on the droplet
surface. These surface charges interact with the applied electric
field and induce fluid motion tangential to the droplet surface.}
\label{fig:ldm}
\end{figure}
As illustrated in Fig.~\ref{fig:ldm}, the interaction of these
surface charges with the applied electric field induces electrokinetic
effects, which ultimately lead to a fluid motion parallel to the
droplet surface. Depending on the ratios of the conductivities,
viscosities, and dielectric constants of both media, the flow
above and below the droplet is directed outwards or inwards. Outward
(inward) flow leads to an oblate (prolate) deformation of the
droplet. The deformation consists of a stationary
and a time-dependent part. The amplitude of the latter turned out to
be below the detection limit of our set-up, \ie is negligible in our
system. In the notation adopted by 
Saville and coworkers \cite{vizika1992,saville1997}, the stationary
deformation of the droplet $D_s = (d_1-d_2)/(d_1 + d_2)$ (with $d_1$ and
$d_2$ being the principal diameters of the droplet) can be expressed by
\begin{equation}
D_{stat} = \frac{9}{16}\epsilon_0 \epsilon_m \frac{a E^2} {\gamma} \Phi,
\end{equation}
with $a$ representing the droplet radius in the absence of an electric
field and $E$ the actual field in the
sample. Since the interfacial tensions $\gamma$ enters in the
denominator, it is clear that the extremely small value of $\gamma
\approx 1.5 \cdot 10^{-5}$ N/m enables large deformation of the
droplets. 
The function $\Phi$ determines the type of the deformation. It is
positive (negative) for prolate (oblate) droplets 
\begin{widetext}
\begin{equation}
\Phi = 1 - \frac{ S^2 R (11+14M) + 15S^2(1+M) + S(19+16M)
        + R^2S \tau \omega (M+1)(S+2)}
        {5(M+1)[S^2(2+R)^2 + R^2 \tau^2 \omega^2(1+2S)^2]}.
\label{eq:ldm}
\end{equation}
\end{widetext} 
Here,  $R = \sigma_d/\sigma_m$ denotes the ratio of
the conductivities, $S = \epsilon_m/\epsilon_d$ the ratio of the
dielectric constants, $M = \nu_m/\mu_d$ the ratio of the
viscosities, $\tau = \epsilon\epsilon_0 / \sigma$ the
characteristic time, $\omega$ the angular frequency of the applied
electric field, and indices $d$ and $m$ indicate the properties of
the droplet and the continuous medium, respectively. $\Phi$ is a
monotonic function of $\omega$ which may change sign as a function of
$\omega$, but has no maximum or minimum at finite
$\omega$. To obtain a maximum in the deformation we have to include
the frequency dependent electric field in the sample discussed above. 

\subsection{Comparison with our data}
\label{sec:comparison-with-our} 

\begin{figure} 
\includegraphics[width=0.45\textwidth]{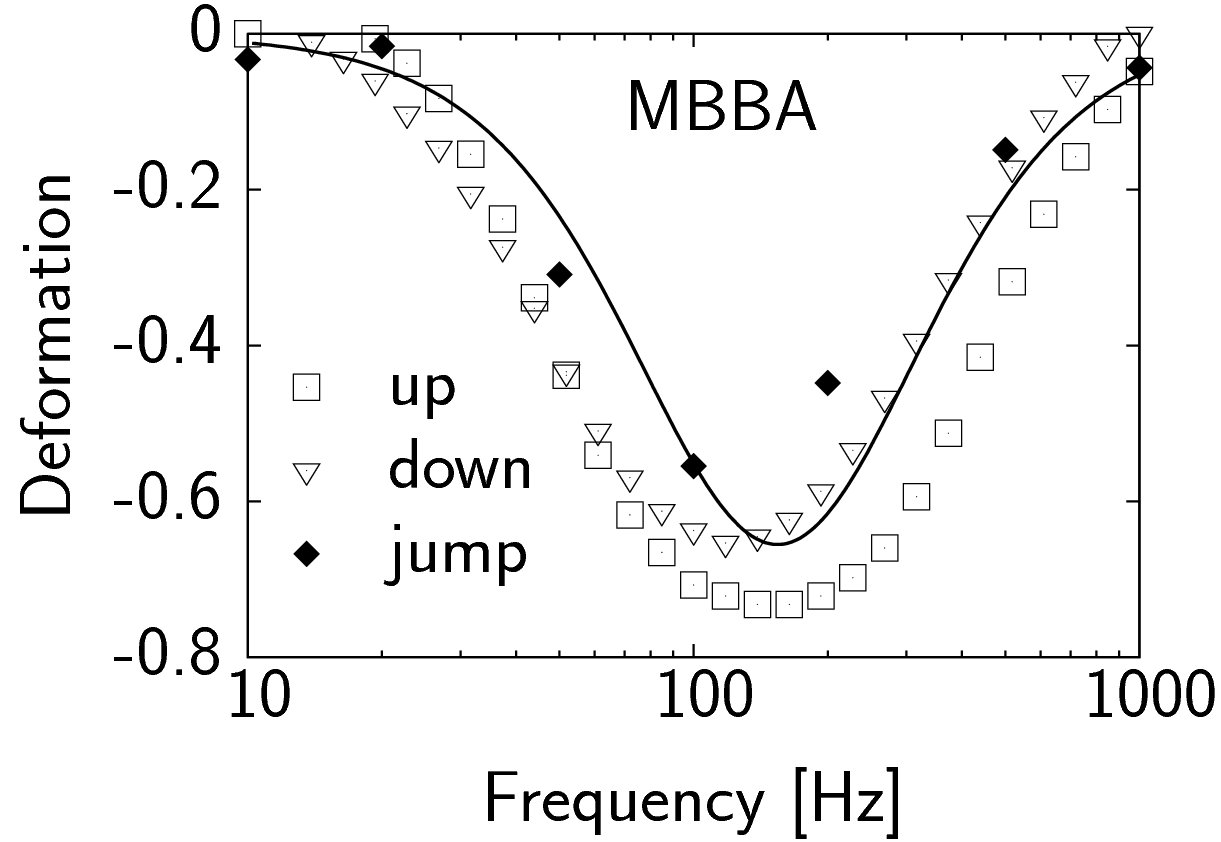}
\caption{The deformation $D_s = (d_1-d_2)/(d_1 + d_2)$ (with $d_1$ and
$d_2$ being the principal diameters of the droplet) as calculated from
the data of Fig.~\ref{fig:mbba-amplitude} (assuming a constant droplet
volume) is compared to  the combination of 
screening and the leaky dielectric model (solid line). The modelling
curve has been calculated with the material parameters given in
Tab.~\ref{tab:mater-param}, 
an applied ac voltage with 200 V amplitude, a droplet radius of 60
$\mu$m, and an electrode distance of 500 $\mu$m. }
\label{fig:mbba-deform-comp}
\end{figure}

\begin{figure}
\includegraphics[width=0.35\textwidth]{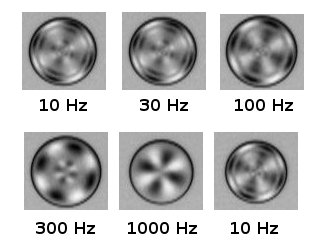}
\caption{Frequency dependent deformation of nematic droplets in a
5CB/ionic liquid mixture (about 0.01 vol\% of 1-butyl-3-methylimidazolium
trifluoromethansulfonimid in 5CB). }
\label{fig:5CB-IL}
\end{figure}

The solid line in Fig.~\ref{fig:mbba-deform-comp} combines our
estimate of the actual voltage applied to the sample,
$E_s = U_s/d$ from Eq.(\ref{eq:actual-field-sample}), with the
frequency dependent deformation given by Eq.~(\ref{eq:ldm}). In this
comparison we used the material parameters (see
Tab.~\ref{tab:mater-param}) given in the literature and chose the
value of the only free parameter in the model, the
conductivity, to fit the measured deformations. The resulting values
for the conductivity ($\sigma_{MBBA} = 5\cdot 10^{-8}$ S/m and
$\sigma_{5CB} = 1.6\cdot 10^{-6}$ S/m) are in the range expected for
liquid crystals. 
With the above assumptions, the combination of leaky dielectric model
and screening due to the finite conductivity accurately describes the
measured data.  

To verify the effect of free charges, we increased the
concentration of free ions by mixing one of the samples (5CB) with
an ionic liquid (1-butyl-3-methylimidazolium
trifluoromethansulfonimide). This strongly suppressed all field-induced
effects. At a concentration of a few vol\% neither a 
change in texture nor a deformation was observed. Even at much
lower charge concentrations, both effects were detectable but
partially suppressed compared to pure samples (see
Fig.~\ref{fig:5CB-IL}). We attribute this strong reduction to
the high concentration of free ions which leads to a significantly
higher conductivity and thus to better screening.

\section{Conclusion}

The extremely low interfacial tension at the nematic-isotropic
interface allows for easy deformation of nematic droplets in the
isotropic-nematic coexistence interval. 

The reported frequency-dependent oblate (disc-like) deformation of
nematic droplets induced by an applied electric ac field   
extents the behaviour expected in "leaky"
dielectrics. A part from the anisotropy in the material parameters, the
nematic degrees of freedom seem not to play a significant role in
mechanisms leading to the droplet deformation. Working with insulated
electrodes suppressed electrochemical reactions at the electrodes but
also made it necessary to include screening in the theoretical
description. 
As characteristic for leaky dielectrics, an
interface-driven, frequency-dependent  hydrodynamic flow field around
the droplets was induced by the electric field. The effective field
however was reduced  due to  screening by field-induced charges at the
interface, caused by the finite conductivity of the
sample. Accordingly, combining screening with the "leaky dielectric"
model accurately describes the experimentally observed droplet
deformation.  



\section*{Acknowledgment}

It is pleasure for GKA to acknowledge fruitful discussions with Harald
Pleiner and Patricia E. Cladis. 
We are grateful to Jochen S. Gutmann for synthesis of the ionic liquid
and to  Gabi Sch\"afer and Norbert H\"ohn for their help with the
data treatment.
This work has been supported by the Deutsche Forschungsgemeinschaft through
the SFB TR6 'Physics of Colloidal Dispersions in External Fields'.

\normalem
\bibliographystyle{apsrev}
\bibliography{/home/guenter/Paper/Literatur/litera-2009-02-25}

\end{document}